\documentclass[twocolumn, amssymb, amsmath, showpacs, aps, prb]{revtex4-1} 
\usepackage[dvips]{graphicx, color} 
\begin{document}
\title{True reentry of the glassy state in geometrically frustrated LiCr$_{1-x}$Mn$_x$O$_2$} 
\author{S. Chattopadhyay}
\author{S. Giri} 
\author{S. Majumdar} 
\email{sspsm2@iacs.res.in} 
\affiliation{Department of Solid State Physics, Indian Association for the
Cultivation of Science, 2A \& B Raja S. C. Mullick Road, Kolkata 700 032, INDIA} 
\author{D. Venkateshwarlu}
\author{V. Ganesan}
\affiliation{UGC-DAE Consortium for Scientific Research,
University Campus, Khandwa Road, Indore 452 017, INDIA}

\begin{abstract}  

The development of spin glass like state in a geometrically frustrated (GF) magnet is a matter of great debate. We investigated the effect of  magnetic (Mn) and nonmagnetic (Ga) doping at the Cr site of the layered GF antiferromagnetic compound LiCrO$_2$. 10\% Ga doping at the Cr site does not invoke any metastability typical of a glassy magnetic state. However, similar amount of Mn doping certainly drives the system to a spin glass state which is particularly evident from the relaxation, magnetic memory and heat capacity studies. The onset of glassy state in 10\% Mn doped sample is of reentrant type developing out of higher temperature antiferromagnetic state. The spin glass state in the Mn-doped sample shows a true reentry with the complete disappearance of the antiferromagnetic phase below the spin glass transition. Mn doping at the Cr site can invoke random ferromagnetic Cr-Mn bonds in the otherwise 120$^{\circ}$ antiferromagnetic triangular lattice leading to the non-ergodic spin frozen state. The lack of spin glass state on Ga doping indicates the importance of random ferromagnetic/antiferromagnetic bonds for the glassy ground state in LiCrO$_2$. Spin glass state in GF system has been earlier observed even for small non-magnetic disorder, and our result  indicates that the issue is quite nontrivial and depends strongly on the material system concerned.
 
\end{abstract}

\pacs{75.47.Lx, 75.50.Lk, 75.10.Jm, 75.40.Cx}      

\maketitle
\section{Introduction}

The term `Geometrical Frustration' denotes a novel class of real systems where the arrangements of magnetic ions in the crystal lattice are such that the spins get frustrated in presence of conflicting exchange interactions resulting a complex magnetic state~\cite{gree,rami,har}. On the other hand, `spin glass' (SG) denotes a group of materials having metastable magnetic ground state associated with cooperative spin freezing in random fashion below a characteristic temperature $T_f$. SG  is essentially a manifestation of chemical disorder along with frustration due to random  as well as competing magnetic interactions~\cite{rami3}. Due to such proximity between SG and geometrically frustrated (GF) systems, sincere effort has been made in recent years to address an important issue, {\it i.e.} whether an SG like ground state can be realized in GF systems in presence of quenched disorder. As a result of rigorous investigations, spin freezing has been revealed in quite a few GF systems, such as SrCr$_8$Ga$_4$O$_{19}$, Gd$_3$Ga$_5$O$_{12}$, Y$_2$Mo$_2$O$_7$, Zn$_{1-x}$Cd$_{x}$Cr$_{2}$O$_{4}$, ZnCr$_{2-x}$Ga$_x$O$_{4}$ and so on~\cite{rami2,schi,gin,fio,rat}. Notably, some of these compositions show  SG freezing even in their stoichiometric form where the amount of quenched disorder is supposed to be negligibly small. Thus, the origin of SG state in a GF material is widely debated, and different models have been proposed to account for that~\cite{aa}. Apart from the origin, often the nature of the SG state in a GF material is found to be unusual. For example, the non linear susceptibility analysis on Tb$_2$Mo$_2$O$_7$ shows unconventional SG state in this chemically ordered GF compound~\cite{dksing}. In case of doped GF sample such as  SrCr$_{8.28}$Ga$_{3.72}$O$_{19}$, where sufficient chemical disorder is present, neutron scattering studies indicate that the ground state magnetic excitation deviates significantly from that expected for a conventional SG~\cite{lee}. It is therefore appears that the SG state in a GF magnet is quite intriguing and there remains several open questions to be addressed.

\par
Chemical substitution in a otherwise stoichiometric GF compound can create random exchange interaction through the breakage or alternation of magnetic bonds. It can also affect the magnetic interaction through the generation of random strains on substitution of atoms of different ionic radii~\cite{aa}. The situation can be more complicated if the substituting atoms are  magnetic in nature, as they can introduce additional magnetic interaction in the system. In this work we have investigated the role of chemical substitution (both magnetic and non-magnetic) on the magnetic ground state of LiCrO$_2$.

\begin{figure}[t]
\centering
\includegraphics[width = 8 cm]{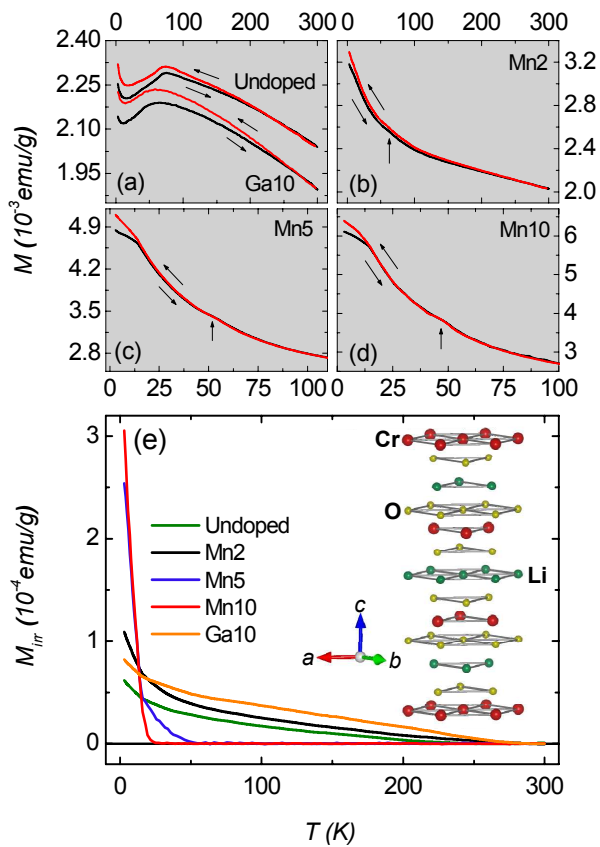}
\caption {Temperature ($T$) variation of magnetization ($M$) measured in zero field cooled (ZFC) and field cooled (FC) protocols with an applied field ($H$) of 100 Oe for (a) undoped and Ga10, (b) Mn2, (c) Mn5, and (d) Mn10 compositions. (e) shows the difference between FC and ZFC magnetization ($M_{irr} = M_{FC} - M_{ZFC}$) as a function of $T$ for all of the compositions. Inset represents a perspective view of the crystal structure of LiCrO$_2$.}
\end{figure}

\par
LiCrO$_2$ is a well known GF magnet having layered hexagonal stacking along the $c$ axis~\cite{lka,sug,seki,ola}. It is a quasi two dimensional magnetic oxide with extremely weak inter-layer interaction. Recently our group has reported  the development of large electric polarization on Cu substitution at the Li site~\cite{giri} which is associated with enormous magneto-structural transition. The Cr$^{3+}$ ions in LiCrO$_2$ remain within edge-sharing distorted CrO$_6$ octahedra and it results a layered arrangements of Cr ions separated by Li$^{+}$ and O$^{2-}$. The magnetic frustration in this compound arises due to triangular arrangements of Cr$^{3+}$ ($S$ = 3/2) ions in the basal plane coupled with each other through AFM type interaction. The compound shows a complex double-$Q$ 120$^{\circ}$ AFM ordering below $T_N$ = 62 K~\cite{kad}, although the Curie-Weiss temperature is substantially high ($\theta_{CW}$ = -700 K)~\cite{ola} resulting a large frustration factor $F$ = $\left|\theta_{CW}\right|$/$T_N$ = 11.3. The 2D triangular network in LiCrO$_2$  provides us a good opportunity to study the effect of chemical substitution on the magnetic ground state. 

\section{Experimental Details}
Polycrystalline samples of LiCr$_{1-x}$Mn$_x$O$_2$ ($x$ = 0.0, 0.02, 0.05, and 0.1, hereafter denoted  by undoped, Mn2, Mn5, and Mn10 respectively) and LiCr$_{1-x}$Ga$_x$O$_2$ ($x$ = 0.1, hereafter denoted by  Ga10 ) were prepared as described elsewhere~\cite{seki}. Powder x-ray diffraction (Cu K$_{\alpha}$) was carried out on all the samples which are found to be single phase in nature having rhombohedral crystal structure (space group: $R\overline{3}m$). The estimated lattice parameters of the undoped sample are $a$ = 2.92 ~\AA ~and $c$ = 14.43 ~\AA, ~which are very close to the previously reported values. DC magnetization ($M$) measurements were performed between 2 and 300 K using a Quantum Design SQUID magnetometer and a cryogen free high magnetic field system from Cryogenic Ltd., U.K. The heat capacity was measured on Quantum Design Physical Properties Measurement System using relaxation technique. Element mapping was performed in a high resolution transmission electron microscope (TEM) from JEOL. 


\begin{table}[t]
\centering
\caption{Variation of the AFM onset temperature ($T_p$), ZFC-FC bifurcation temperature ($T_{irr}$), coercive field ($H_C^{3 K}$) at 3 K, and difference between FC and ZFC magnetization ($M_{irr}^{3K}$) at 3 K for LiCr$_{1-x}$A$_x$O$_{2}$ (A = Ga, Mn).}
\begin{tabular}{ccccc}
\hline
\hline
$Compositions$ & $T_p$ &  $T_{irr}$ & $H_C^{3 K}$ & $M_{irr}^{3K}$  \\ 
 & (K) &  (K) & (Oe) & (10$^{-4}$ emu/g) \\
\hline
Undoped & 75  &  $\sim$250 & -- & 0.6 \\
Ga10 & 65 & $\sim$285  &  --  & 0.8 \\
Mn2 & $\sim$54 & $\sim$280 & -- & 1.1   \\ 
Mn5 & $\sim$51 & $\sim$50 & $\sim$90 & 2.5  \\ 
Mn10 & $\sim$45 & $\sim$20 & $\sim$200 & 3.1 \\ 

\hline
\hline

\end{tabular}
\label{values}
\end{table} 

\section{Results and Discussions}

The $T$ variation of $M$ for doped and undoped samples is depicted in fig. 1. Measurements were performed in zero-field-cooled (ZFC) and field-cooled (FC) conditions under an applied magnetic field of 100 Oe. Both ZFC and FC curves of LiCrO$_2$ increase gradually below 300 K and show a peak around $T_p$ = 75 K which signifies the onset of AFM transition as reported earlier~\cite{lka}. Below 15 K, $M$ shows a sharp upturn with decreasing $T$. Such upturn in low dimensional magnetic systems is often attributed to  paramagnetic impurities and/or broken chain effect~\cite{moto,sc2}. A bifurcation between ZFC and FC curves appears below $T_{irr}$ $\sim$ 250 K for the pure sample. In LiCrO$_2$, such irreversibility was reported earlier and it was attributed to  the formation of {\it frozen magnetic clusters} having short range magnetic correlations~\cite{lka}. On Ga doping (see the $M$-$T$ curve of Ga10), the overall behavior of the thermomagnetic curve remains almost unaltered. Although, the peak becomes broader with a lowering of $T_p$ (65 K). The magnitude of $M$ decreases slightly in case of Ga-doped sample.

\par
In contrary, Mn doping at the Cr site causes considerable change in the magnetic properties of LiCrO$_2$.  Figs. 1(b)-1(d) show ZFC and FC $M$($T$) curves of Mn2, Mn5, and Mn10 samples respectively, between 2 K and 300 K.  A relatively weak hump like feature in all the doped samples can be observed in the ZFC or FC curves signifying the onset of the AFM transition. The anomaly is found to be around $T_p$ $\sim$ 54 K, 51 K and 45 K for Mn2, Mn5, and Mn10 samples respectively. In Mn5 and Mn10, the low-$T$ {\it Curie tail}-like  rise is completely absent as well. The bifurcation point between FC and ZFC data  gradually shifts to lower $T$ with increasing Mn concentration (see table I). We have plotted $M_{irr} = M_{FC}-M_{ZFC}$ in fig. 1(e) to have a quantitative idea of the irreversibility in $M$. For undoped, Ga10 and Mn2 samples $M_{irr}$ becomes non-zero below about 250 K, 285 K and 280 K respectively, whereas for Mn5 and Mn10 samples, it is negligibly small down to 50 K and 20 K respectively. However, below this temperatures $M_{irr}$ rises sharply for Mn5 and Mn10 indicating large thermomagnetic irreversibility (see table I). Such low temperature rise in $ M_{irr}$ is absent in other samples and it signifies a different mechanism for the observed irreversibility in these two compositions.

\begin{figure}[t]
\centering
\includegraphics[width = 8.5 cm]{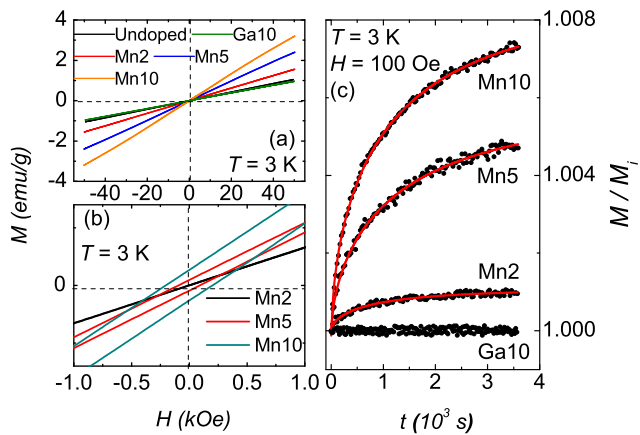}
\caption {(a) shows isothermal  magnetization ($M$) as a function of applied magnetic field ($H$) for all the compositions at 3 K. (b) emphasizes the low field regime of $M$-$H$ curves of Mn doped samples to depict the compositional variation of coercivity. (c) represents the time ($t$) variation of normalized $M$ for Ga and Mn doped samples measured in zero field cooled condition at 3 K under an applied field of 100 Oe. Here, $M_i$ denotes the initial magnetization in the beginning of measurement and solid lines are fit to the relaxation data with equation 1.}
\end{figure}
\begin{figure}[t]
\centering
\includegraphics[width = 8 cm]{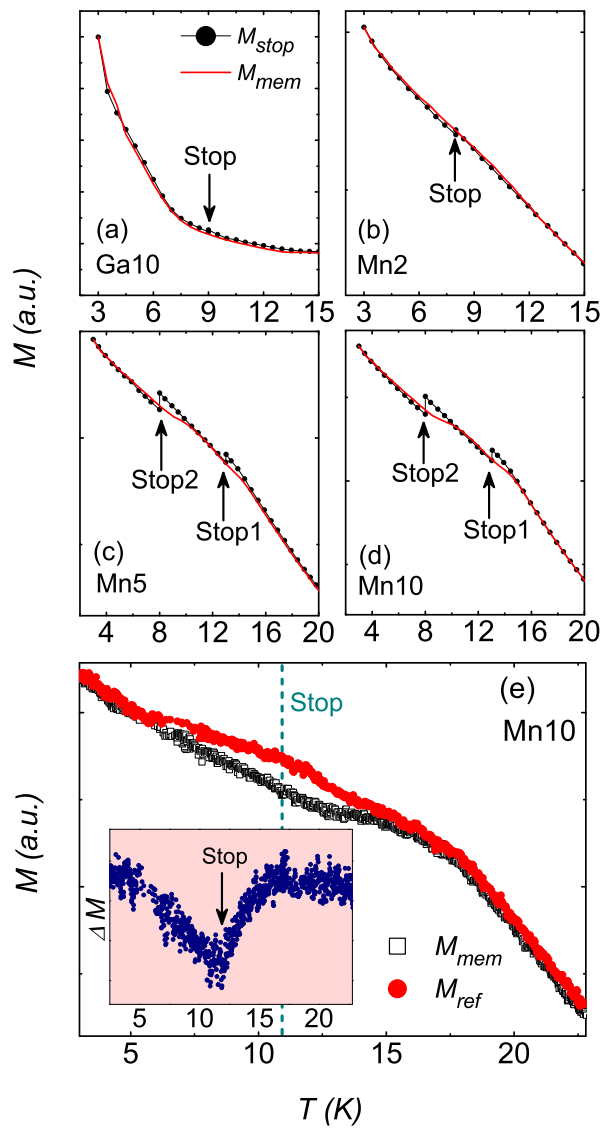}
\caption {(a)-(d) depict memory measurements following field stop field cooled protocol for Ga10, Mn2, Mn5, and Mn10 respectively with an applied field of 100 Oe. Here, $M_{stop}$ denotes the cooling curve with stops of 3600 s each, and $M_{mem}$ is the subsequent heating curve without any stop. (e) shows the ZFC memory measurement on Mn10. Where, $M_{mem}$ represents the ZFC heating curve after the sample being cooled in zero field with an intermediate stop at 12 K for 18000 s. $M_{ref}$ denotes the reference ZFC curve where the data was recorded during heating  after the sample being cooled without any stop. Inset shows $T$ dependence of the difference curve $\Delta M = M_{mem} - M_{ref}$ to illustrate the ZFC memory effect.}
\end{figure}

\par
Fig. 2(a) depicts isothermal field dependence of $M$ at 3 K for all the samples. For the undoped and Ga10 samples, the $M$-$H$ data is linear and does not show any hysteresis. This signifies strong AFM character. $M$($H$) of the Mn-doped samples show slight non-linearity especially at the high field region with curvature increasing with increasing Mn concentration. The magnitude of $M$ is found to rise with Mn doping, while for 10\% Ga doped sample it is almost same as of the undoped sample. Fig.2(b) emphasizes the low field region of $M$($H$) curves of the Mn doped samples. Interestingly, the presence of hysteresis can be observed in Mn5 and Mn10 compositions with a systematic enhancement of coercivity with Mn content (see table I). 

\par
An SG like sate is generally characterized by a free energy landscape with innumerable numbers of nearly degenerate ground state configurations separated from each other by potential barriers of random height (landscape of random potential wells)~\cite{binder}. A signature of such scenario is the presence of  magnetic relaxation, which occurs due to the passage from one metastable state to another with time. Fig.2(c) depicts normalized $M$ {\it vs.} time($t$) data for the Mn and Ga doped samples at 3 K. To record this data, samples were first zero field cooled from 300 K, and a field of 100 Oe was switched on. All the Mn-doped samples show the presence of finite relaxation with the magnitude getting higher with Mn concentration. However, we failed to observe any measurable relaxation in undoped and Ga10 sample. These relaxation data in Mn-doped samples are found to be best described by a modified stretched exponential law~\cite{myd,phi,rvc,fre},

\begin{equation}
M(t) = M_i - M_{r}exp[-(t/\tau)^\beta]
\label{expo} 
\end{equation}

where, $M_i$ is the initial magnetization, $M_{r}$ is the amplitude of the  metastable part, $\tau$ is the time constant and the parameter $0 \leq \beta \leq 1$ signifies the distribution of local energy minima. It  approaches unity  for a system with long range magnetic order. Fitting to the $M$($t$) data with this equation results $\beta$ = 0.64, 0.63, 0.60 for Mn2, Mn5, and Mn10 respectively. The gradual decrease of the value of $\beta$ signifies that the system passes through increased number of local minima in the energy landscape during relaxation. Systems having spin glass like ground states are found to show $\beta$ ranging between $\sim$0.2-0.6~\cite{wan,chu,bha,cha2}. The estimated $\beta$ values of the present compositions practically fall in this regime.

\par
Evidently,  the signature of metastability in Mn -doped samples are not visible in the undoped or Ga-doped samples. 
It thus seems that the magnetic ground state in nonmagnetic Ga-doped sample is different from that of magnetic Mn-doped counterparts. A possible way to distinguish a glassy metastable state from a long range ordered one is through magnetic memory measurements. We performed field stop field cool (FSFC)  memory in $M$($T$) measurement on the Mn and Ga doped samples (see fig.3)~\cite{sas,sun}. In this protocol, sample was first field cooled from 300 K down to 5 K with intermediate stops of 3600 s each with the magnetic field being reduced to zero during each stop (denoted as $M_{stop}$). After cooling, the samples were subsequently heated back to 300 K without any stop in presence of $H$ (denoted as $M_{mem}$). A system with frozen/blocked spins (or spin clusters) is expected to show anomalies in the heating curve at the same temperatures where the sample was allowed to age during cooling. We do not observe any such anomalies in Ga10 and Mn2 samples (figs. 3 (a) and (b) respectively) at the stopping temperatures. On the other hand $M_{mem}$ curves for Mn5 and Mn10 (figs. 3(c) and 3(d) respectively) show anomalies in $M_{mem}$ with clear  changes in slope at the stopping positions. The anomalies associated with the memory effect is found to be much stronger in Mn10 sample.

\begin{figure}[t]
\centering
\includegraphics[width = 8 cm]{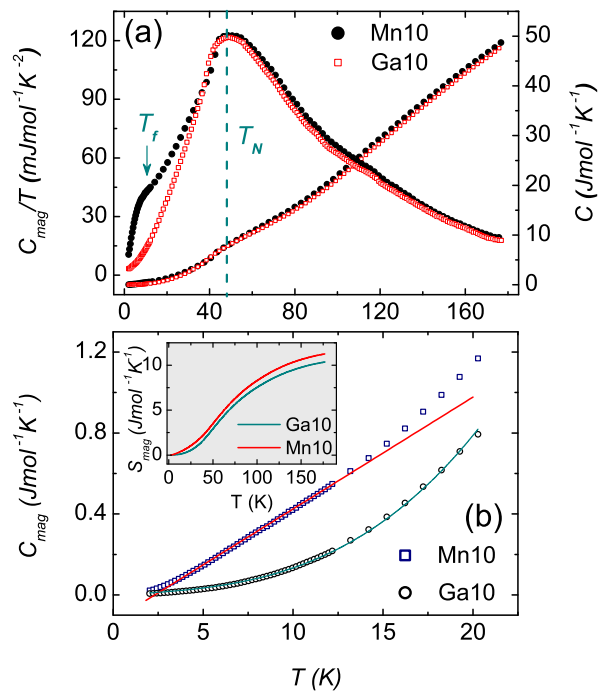}
\caption {(a) Right axis depicts temperature variation of the heat capacity ($C$) between 2 K and 180 K for Mn10 and Ga10 compositions. Whereas, the left axis represents variation of $C_{mag}/T$ with temperature. Here, $C_{mag}$ is the magnetic contribution of heat capacity. (b) shows the low-$T$ side of $C_{mag}$ for Mn10 and Ga10 as a function of $T$. Solid lines are the best fit to the data with appropriate algebraic expressions (see text for details). Inset shows $T$ variation of magnetic entropy ($S_{mag}$) of both the samples.}
\end{figure}

\par
Although the existence of FSFC memory effect is a convincing signature of a frozen or blocked  magnetic state, it cannot distinguish between a spin glass and superparamagnet. In order to resolve this issue,  memory measurement in ZFC condition was performed on Mn10. Here the protocol remains almost the same as of FSSC baring the fact that the cooling was performed in zero field with a stop for 18000 s at 12 K. Fig. 3 (e) shows the subsequent heating curves ($M_{mem}$) along with the reference curves ($M_{ref}$), which is actually simple ZFC heating curve without any stop during cooling. The difference curve $\Delta M$ = $M_{mem}$ - $M_{ref}$ is plotted for  Mn10 in the insets of fig. 3(e).  Interestingly, $M_{mem}$ and $M_{ref}$ curves for Mn10 follows different path around the stopping temperature of 12 K. This feature gets much more prominent in the difference curve as shown in the inset of fig. 3(e), where a  dip can be seen with its minimum at $T_{stop}$ = 12 K. The appearance of ZFC memory in Mn10 signifies that the substitution of Cr ions with Mn disrupts the AFM order and turns on a spin glass like magnetic ground state. 

\begin{figure*}[t]
\centering
\includegraphics[width = 14 cm]{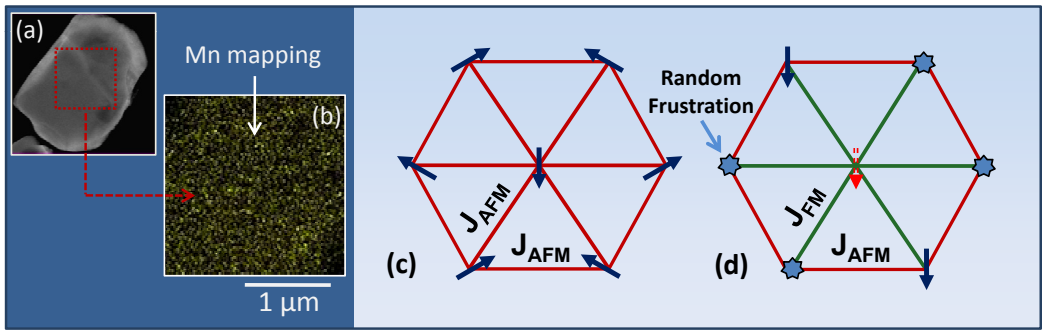}
\caption {(a) shows TEM image of an Mn10 crystallite. (b) illustrates  element mapping on the selected region of the same particle using high resolution transmission electron microscope with bright spots denoting Mn. (c) represents cartoon of a frustrated triangular latice with nearest neighbor AFM interaction and 120$^\circ$ noncolinear AFM order. (d) depicts how the generation of random FM bonds by means of doping (dashed spin) can destroy the ordering.}
\end{figure*}

\par
It is now pertinent to know the nature and origin of SG state in Mn10 sample, particularly to what extent it corresponds to a conventional SG state. We addressed this point through heat capacity ($C$) measurements on both Mn10 and Ga10 samples as depicted in fig. 4. In the $C$ versus $T$ data [see fig. 4(a)], a clear anomaly around 45 K is observed in both the samples which signifies the long range AFM transition. Apparently the $C$ vs. $T$ data of two samples look identical, however a careful look brings out some subtle differences which are important to characterize the magnetic ground state of these compositions. The magnetic contribution to the specific heat ($C_{mag}$) has been calculated  by subtracting the lattice contribution ($C_{latt}$) from total heat capacity $C$. Here $C_{latt}$ has been estimated from the heat capacity data of isostructural non-magnetic compound LiCoO$_2$ taken from reference~\cite{licoo2} followed by proper scaling as prescribed by Bouvier {\it et al}~\cite{bouv}. The main panel of fig. 4(a) (left axis) shows the $C_{mag}/T$ versus $T$ plot. $T_N$ in both the samples are  seen as a broad peak around 45 K (this resembles well with the  reported $C_{mag}/T$ versus $T$ data of pure LiCrO$_2$~\cite{lka}). A marked difference is observed between $C_{mag}/T$ data of Mn10 and Ga10 below $T_N$. For Ga10, $C_{mag}/T$ smoothly decreases with decreasing $T$, whilst  $C_{mag}/T$ shows a shoulder like feature below about 14 K in Mn10. This feature matches well with the FC-ZFC  bifurcation temperature $T_{irr}$ in the $M(T)$data of the same sample.

\par
We have carefully examined the low temperature part of $C_{mag}$ of both the samples as depicted in fig. 4(b). A higher value of $C_{mag}$ is observed for the Mn10 sample. In Mn10, the larger value of the magnetic entropy ($S_{mag}$) calculated from $C_{mag}$ (see inset of fig. 4(b)) is  related to the excess magnetic disorder. For an AFM sample with linear $\omega-q$ dispersion relation of magnons, one expects $C_{mag}$ to vary as $T^3$ at low-$T$. A predominant $T^3$ dependent $C_{mag}$ is found for Ga10, which signifies an ordered AFM state. We have fitted the observed $C_{mag}$ data of Mn10 and Ga10 samples using the relevant algebraic expressions and it brings out some novel information regarding their respective magnetic ground states. For Ga10, it is found that a simple $T^3$ is unable to provide a good fit to $C_{mag}$ below 14 K. Rather an equation  $C_{mag} = a_1T + a_3T^3$ provides the best fit to the data with $a_1$ = 3.6 mJ(mol)$^{-1}$K$^{-2} $ and $a_3$ = 9.77$\times$10$^{-2}$ mJ(mol)$^{-1}$K$^{-4}$. Since these samples are highly insulating, an electronic origin of linear $T$ term in $C_{mag}$ can be ruled out. Such linear $T$ term is likely to have magnetic origin and probably it denotes some disordered magnetic state.

\par
Now turning to Mn10, $C_{mag}$ shows a linear behavior below 14 K, which  corresponds an SG state (since the sample is an insulator)~\cite{martin}. An upward curvature is observed below about 3 K in the $C_{mag}(T)$ data of Mn10. The linear part between 12 and 3 K, if extrapolated, hits the abscissa at a positive finite value of $T$. Such finite intercept and low-$T$ curvature are previously observed in several canonical SGs~\cite{binder,martin}. $C_{mag}$ between 12 and 3 K was fitted by an algebraic formula $C_{mag} = b_0 + b_1T$ with $b_0$ = -127.66 mJ(mol)$^{-1}$K$^{-1}$ and $b_1$ = 55.25 mJ(mol)$^{-1}$K$^{-2}$. 

\par
The SG state in Mn10 sample is of reentrant type, which develops out of an AFM ordered state. Occurrence of reentrant spin glass (RSG) in AFM system, albeit fewer in number than its FM counterpart, are well documented in the literature~\cite{dho,chen,zha,nim}. It has been argued  that Mn$^{3+}$-Cr$^{3+}$ interaction can be FM type particularly when the metal atoms are in octahedral oxygen environment~\cite{jon,su,xia,dei,gan}. Substitution of Mn at the Cr site can give rise to random FM bonds in the otherwise 120$^{\circ}$ non-co-linear AFM structure of LiCrO$_2$ (see fig. 5). In the random field model of RSG as proposed by Aeppli {\it et al.},~\cite{aep} presence of spatially uncorrelated conflicting magnetic bonds can destroy the long range FM ordering through the emergence of random microscopic field. The system then attains a frozen non-ergodic state similar to conventional spin glasses. An equivalent scenario can occur when an AFM ordering is destroyed resulting a reentrant glassy state, which would be appropriate for the present Mn-doped LiCrO$_2$ sample. Similar argument was actually provided in case of AFM spin glass observed in Cr doped manganites~\cite{dho}.

\par
A cartoon of the 120$^{\circ}$ AFM spin arrangement on a 2D hexagonal lattice is depicted in fig. 5(c). Substitution of one magnetic impurity (dashed spin in fig. 5(d)) at the center of the hexagon with nearest neighbor FM interaction (similar to Cr-Mn FM bond in Mn doped LiCrO$_2$) can destroy the ordering arrangement through the development of conflicting sense of interactions. If such substitutions are spatially random, the system will never attain an order arrangements of spins (like the regular GF system without quenched disorder) and may lead to a glassy state.

\par
It has long been argued whether the ground state of an RSG system is truly spin glass like or SG state coexists with the long range ordered state~\cite{kj,mir,kim,nii}. Unlike Ga10, we do not observe any $T^3$ term in the $C_{mag}$ versus $T$ (expected for an AFM state) plot below 14 K of Mn10. This indicates that the ground state of Mn10 is truly SG type without any coexisting AFM component. Doping of magnetic atoms often results  superparamagnetic like phase due to the clustering of dopant atoms (as in Cu-Mn or Au-Fe alloys~\cite{myd}). In that case one would expect a $T^{3/2}$ term in the $C_{mag}$ versus $T$ plot due to intra-cluster FM coupling.~\cite{thomp} A clear linear variation of $C_{mag}$ in Mn10 ruled out the possibility of any superparamagnetic clustering effect. We  physically examined the sample using high resolution transmission electron microscope for element mapping, where the spatial distribution of the elements present in the sample can be observed. Our data (see fig. 5) does not indicate any clustering of Mn atoms even down to the nanometer scale. It is generally believed that a true RSG state exists only in a magnetic system below three dimension~\cite{kj} and thus Mn10 can be regarded as a novel example of such complete reentry for its layered quasi-2D structure.

\section{Conclusion}

We would like to conclude by comparing our result with the recently published work on doped ZnCr$_2$O$_4$, where a quasi-spin glass state is observed on small amount (1\% to 5\%) of Ga doping~\cite{rami-zco}. The quasi spin glass state is characterized by thermomagnetic irreversibility and a linear term in the $C$ versus $T$ data. The authors argued that the substitution of Ga at the Cr site creates {\it quasi-spins} which give rise to  some degree of non-ergodicity. For our case even 10\% Ga doping does not give rise to a true spin glass state. However, it should be noted that a linear-$T$ part is indeed present on top of the cubic part in $C_{mag}$ of Ga10. The coefficient of linear term is 3.6 mJ(mol)$^{-1}$K$^{-2}$ which is substantially smaller than the linear term in Ga doped ZnCr$_2$O$_4$ (9.6 and 47.6 mJ(mol)$^{-1}$K$^{-2}$ on 1\% and 5\% Ga doping respectively). Ga10 is certainly not a spin glass, but it resembles to some extent with the quasi spin glass state reported in doped ZnCr$_2$O$_4$. Both  LiCrO$_2$ and  ZnCr$_2$O$_4$ are GF magnetic systems, although they are quite different as far as the  crystal structure (layered and cubic spinel respectively) and dimension of magnetic interaction (quasi-2D and 3D respectively) are concerned. ZnCr$_2$O$_4$  undergoes first order phase transition paving the path for disorder through strain field distribution~\cite{kag,lee2}. These factor may enhances the chances of added metastability in case of Ga doped ZnCr$_2$O$_4$. 

\par
Summarizing, a true SG state is only observed in LiCrO$_2$ on Mn doping at Cr site, whilst similar Ga doping does not lead to an SG state. The 10\% Mn-doped sample shows a true reentrance of SG phase in an otherwise AFM state. It appears that random magnetic impurities are essential for the development of an glassy magnetic state in this GF compound. This is quite different from the case of other GF systems where very small non-magnetic impurity can turn the system to a glassy phase. Therefore, it seems that there is no generalized rule  for the development of glassy state in a GF system.

\par 
SC wishes to thank CSIR (India) for his research studentship.  The funding from  CSIR (grant number: 03(1209)/12/EMR-II) for the present work is thankfully acknowledged. The authors also like to thank  the Low Temperature \& High Magnetic Field (LTHM) facilities at UGC-DAE CSR, Indore (sponsored by DST) for heat capacity measurements.

\end{document}